\documentstyle[psfig,prl,multicol,aps]{revtex}
\tighten
\begin{document}
\preprint{\large\it  PC-49 EP2DS-13, Ottawa 1999}
\title{
Two-dimensional charged magneto-excitons: \\
Magnetic translations and localization}
\author{A. B. Dzyubenko\cite{ABD}}
\address{
 Institut f\"{u}r Theoretische Physik, J.W. Goethe-Universit\"{a}t,
 60054 Frankfurt,  Germany }
\author{A. Yu. Sivachenko}
\address{The Weizmann Institute of Science, Rehovot 76100, Israel }
\date{\today}
\maketitle
\begin{abstract}
For composite complexes --- quasi-two-dimensional (2D) charged
magneto-excitons $X^-$ --- we propose a new exact
classification of states,
which is based on magnetic translations.
We consider implications of this symmetry for magneto-optical
transitions. It is shown, in particular,
that in a translationally-invariant system with a simple valence band,
the ground triplet $X^-_t$ state is dark in interband transitions at
finite magnetic fields $B$.
This exact result calls for the re-interpretation of several
previous theoretical studies.
We consider the symmetry-breaking effects of spatial lateral
confinement on internal transitions of charged complexes.
\end{abstract}


\parskip0pt

\begin{multicols}{2}

Recently, there has been considerable experimental and theoretical
interest in charged excitons $X^-$ and $X^+$ in magnetic fields $B$
in 2D systems.
Experimentally, magneto-optical interband \cite{inter} and, more recently,
intraband internal \cite{EP2DS} transitions of charged excitons
have been studied. Theoretically, the binding of
charged excitons $X^-$ has been considered in quantum dots \cite{Hawr,Chap},
in a strictly-2D system in the high-magnetic field limit \cite{AHM},
and in realistic quantum wells (QW's) at finite $B$ \cite{Whit97}.
In all these theoretical works on charged excitons,
the existing exact symmetry --- magnetic translations ---
has not been identified. Some implications of this
symmetry for internal transitions of charged complexes
have been indicated recently in \cite{EP2DS,Dz&S_pr}.
The aim of the present work is to establish and describe in some
detail the underlying hidden symmetry and reveal its striking manifestations
in interband and intraband magneto-optical transitions.

We consider a system of interacting particles of charges $e_j$
in a magnetic field ${\bf B}=(0,0,B)$ described by the Hamiltonian
\begin{equation}
                \label{H} 
   H = \sum_j \frac{ \hat{ \bbox{ \pi }}_j^2}{2m_j}
          + \case{1}{2} \sum_{i \ne j} U_{i j}({\bf r}_i-{\bf r}_j) \, ,
\end{equation}
here
$\hat{\bbox{\pi}}_j = -i\hbar \bbox{\nabla }_j -
\frac{e_j}{c} {\bf A}({\bf r}_j)$
is the kinematic momentum operator of the $j$-th particle in ${\bf B}$ and
$U_{ij}$ are the potentials of interactions that can be rather arbitrary.
Dynamical symmetries of (\ref{H}) are the following.
In the symmetric gauge ${\bf A} = \frac12 {\bf B} \times {\bf r}$,
there is the axial symmetry about the $z$-axis
$[H, \hat{L}_z]=0$, where
$\hat{L}_z=\sum_j ({\bf r}_j \times -i\hbar\bbox{\nabla }_j)_z$.
Therefore, the total angular momentum projection $M_z$,
an eigenvalue of $\hat{L}_z$, is a good quantum number.
In a uniform ${\bf B}$, the Hamiltonian (\ref{H}) is also invariant under
a group of magnetic translations whose generators are the components
of the operator $\hat{\bf K} = \sum_{j} \hat{\bf K}_j$,
where $\hat{\bf K}_j =
\hat{\bbox{ \pi }}_j - \frac{e_j}{c} {\bf r}_j \times {\bf B}$
(see, e.g., \cite{Simon}).
$\hat{\bf K}$ is the exact integral of the motion: $[H, \hat{\bf K}]=0$.
The components of $\hat{\bf K}$ and
$\hat{\bbox{ \pi }} = \sum_j \hat{\bbox{ \pi }}_j $ commute in
${\bf B}$ as
\begin{equation}
	\label{comK}
 [\hat{K}_x, \hat{K}_y] = - [\hat{\pi}_x, \hat{\pi}_y] =
    - i \frac{\hbar B}{c} Q \quad , \quad
 \quad Q \equiv \sum_j e_j  \, ,
\end{equation}
while $[\hat{K}_p, \hat{\pi}_q]= 0$, $p,q=x,y$.
For neutral complexes (atoms, excitons, biexcitons) $Q=0$,
and classification of states in $B$ are due to the continuous two-component
vector --- the 2D magnetic momentum ${\bf K}= (K_x,K_y)$.
For charged systems the components of $\hat{\bf K}$
cannot be observed simultaneously.
This determines the macroscopic
Landau degeneracy of exact eigenstates of (\ref{H}).
For a dimensionless operator
$\hat{{\bf k}} = \sqrt{c/2 \hbar B |Q|} \, \hat{\bf K}$
we have $[\hat{k}_x, \hat{k}_y]=-iQ/|Q|$. Therefore,
$\hat{k}_{\pm}= (\hat{k}_x  \pm i \hat{k}_y)/\sqrt{2}$
are Bose raising and lowering ladder operators:
$[\hat{k}_{+}, \hat{k}_{-}]=-Q/|Q|$.  It follows then that
$\hat{{\bf k}}^2 =
\hat{k}_{+} \hat{k}_{-} + \hat{k}_{-} \hat{k}_{+}$
has discrete oscillator eigenvalues $2k+1$, $k=0, 1, \ldots$.
Since $[\hat{{\bf k}}^2, H]=0$ and $[\hat{{\bf k}}^2, \hat{L}_z]=0$,
the exact charged eigenstates of (\ref{H}), in addition
to the electron $S_e$ and hole $S_h$ spin quantum numbers,
can be simultaneously labeled
by the discrete quantum numbers $k$ and $M_z$.
The labelling therefore is $|k M_z S_e S_h \nu \rangle$,
where $\nu$ is a ``principal'' quantum number, which
can be discrete (bound states) or continuous (unbound states forming
a continuum) \cite{Dz&S_pr}.
The $k=0$ states are {\em Parent States} (PS's)
within a degenerate manifold.
All other {\em daughter} states in each $\nu$-th family
are generated out of the PS iteratively: for $Q<0$
$|k,M_z-k,S_e S_h \nu \rangle =
(\hat{k}_{-})^k |0, M_z, S_e S_h \nu \rangle / \sqrt{k!}$.

It is interesting to consider the following question:
is it possible to construct a complete orthonormal
basis with simultaneously fixed quantum numbers $M_z$ and $k$
(and $S_e$, $S_h$)?
The answer is positive. Let us demonstrate this on the
example of the charged exciton $X^-$.
The basis for the $X^-$ with fixed values
of $M_z$ and $S_e$ is of the form \cite{Dz_PLA,EP2DS}
$\phi^{(e)}_{n_1 m_1}({\bf r}_e) \,
 \phi^{(e)}_{n_2 m_2}({\bf R}_e) \,
 \phi^{(h)}_{N_{h}M_{h}}({\bf r}_{h})$
and includes different three-particle $2e$--$h$ states
such that the total angular momentum projection
$M_z= n_1 + n_2 -m_1 -m_2 - N_h + M_h$ is fixed.
Here $\phi^{(e,h)}_{n m}$ are the $e$- and $h$- single-particle
factored wave functions in $B$, $n$ is the Landau-level (LL) quantum number,
and $m$ the oscillator quantum number [$m_{ze(h)}= {+ \atop (-)}(n-m)$].
We use the electron relative and center-of-mass (CM) coordinates:
${\bf r}_e = ({\bf r}_{e1} - {\bf r}_{e2})/\sqrt{2}$ and
${\bf R}_e = ({\bf r}_{e1} + {\bf r}_{e2})/\sqrt{2}$.
Permutational symmetry requires that for electrons
in the spin-singlet $S_e=1$ (triplet $S_h=1$) state
the relative motion angular momentum $n_1-m_1$ should be even (odd).
This basis complies with
the axial symmetry about the $z$-axis and the permutational symmetry.
To make it compatible with magnetic translations
we shall perform a unitary transformation.
To this end we introduce the intra-LL ladder operators
$B^{\dag}_{e}$ and $B^{\dag}_{h}$ (see, e.g., \cite{Dz_PLA}),
corresponding to the ${\bf R}_e$ and ${\bf r}_h$ degrees of
freedom, respectively, and perform Bogoliubov canonical transformations
$B_e^{\dag} \rightarrow \tilde{B}_e^{\dag} = S B_e^{\dag} S^{\dag}=
u B^{\dag}_{e} + v B_{h}$,
$B_h^{\dag} \rightarrow \tilde{B}_h^{\dag} = S B_h^{\dag} S^{\dag}=
u B^{\dag}_{h} + v B_{e}$,
where $S= \exp \{ \Theta (B_eB_h - B^{\dag}_{h} B^{\dag}_{e} ) \}$
and $u= {\rm ch} \Theta$, $v= {\rm sh} \Theta$.
Choosing ${\rm th} \Theta = 1/\sqrt{2}$
we make $\hat{{\bf k}}^2 = 2\tilde{B}_e^{\dag} \tilde{B}_e  + 1$
diagonal in $\tilde{B}_e^{\dag}$. Therefore, using
transformed operators $\tilde{B}_e^{\dag}$ and $\tilde{B}_h^{\dag}$
operating on the new vacuum
$|\tilde{0} \rangle = S |0\rangle =
({\rm ch}\Theta)^{-1}
\exp\{ - {\rm th}\Theta \, B^{\dag}_{h} B^{\dag}_{e} \} |0\rangle$
solves this problem.
For a system of $N_e$ electrons and $N_h$ holes (with, e.g., $N_e>N_h$),
the analogous transformation should involve the
intra-LL $e$- and $h$- CM operators $B^{\dag}_{e}$ and $B_{h}$
with ${\rm th} \Theta = \sqrt{N_h/N_e}$.

Let us discuss now magneto-optical transitions of charged
complexes.
In the dipole approximation the photon momentum is negligibly small.
Therefore, the quantum number $k$ should be conserved in
intra- \cite{EP2DS,Dz&S_pr} and inter-band magneto-optical transitions.
Let us establish this selection rule more formally.
For internal {\em intraband} transitions
the Hamiltonian of the interaction with the
radiation of polarization $\sigma^{\pm}$ is of the form
$\hat{V}^{\pm} \sim  \sum_{j} e_j \pi_{j}^{\pm}/m_j$,
where $\pi_j^{\pm} = \pi_{jx} \pm i \pi_{jy}$.
Conservation of $k$ follows from the commutativity
$[\hat{V}^{\pm}, \hat{\bf K}]=0$.
Other selection rules in this case
are conservation of spins and
$\Delta M_z= \pm 1$ in the $\sigma^{\pm}$ polarization
for the envelope function.
For {\em interband} transitions the interaction with the radiation
field is described by the luminescence operator
$\hat{\cal L}= p_{\rm cv} \int \! d{\bf r} \,
\hat{\Psi}^{\dagger}_{e}({\bf r}) \hat{\Psi}^{\dagger}_{h}({\bf r})
+ \mbox{H.c.}$, where $p_{\rm cv}$ is the momentum interband
matrix element.
Conservation of $k$ follows from $[\hat{\cal L},\hat{\bf K}]=0$.
Other selection rules for this case are conservation of spins and
$\Delta M_z= 0$ in the $\sigma^{\pm}$ polarization
for the envelope function
(the Bloch parts change according to $\Delta m_z= \pm 1$).
Conservation of $k$ constitutes a new exact selection rule.
It can be formulated in a simple verbal way: {\em Parents
can talk only to parents, while daughters talk to daughters
if and only if their parents do talk.}

Let us consider implications of this selection rule for
the interband transition --- photoluminescence from the triplet ground state
$X^-_t \rightarrow e^-_{n}+ h\nu$ with the emission of the photon
and the electron  in the $n$-th LL in the final state.
It turns out that simultaneous conservation of $M_z$ and $k$
in this transition makes it strictly prohibited:
It is now well established \cite{AHM,Whit97,Dz&S_pr}
that the triplet $X^-_{t}$ ground PS with $k=0$ has $M_z=-1$
at finite $B>8$\,T and in quasi-2D QW's.
On the other hand, an electron in the $n$-th LL, $e^-_{n}$,
has $M_z=n-k$ (the oscillator quantum number $m=k$).
We see that $\Delta k=0$, $\Delta M_z=0$ cannot be satisfied
for transition to any electron LL $n$ if the $X^-$ PS has
$M_z <0$.
This means that the ground triplet state $X^-_t$ is optically {\em dark}
in interband transitions (note that the singlet ground state $X^-_s$ has
$M_z=0$ and is optically active).
In the 2D high-$B$ limit this also follows \cite{AHM}
from the so called ``hidden symmetry'' in $e$--$h$ systems \cite{hidden}.
We stress that our result is much more general.
Indeed, quasi-2D effects, $e$--$h$ asymmetry, admixture of
higher LL's (finite $B$-effects) break neither axial
nor translational symmetry. Therefore,
even in the presence of these effects, the triplet stays dark ---
as long as the ground $X^-_t$ PS has $M_z<0$.
This exact result calls for re-interpretation of
several previous theoretical works on optical
properties of the triplet $X^-_t$ ground state \cite{AHM,Chap,Whit97}.
Very small but finite oscillator strengths of the
$X^-_t$ ground state obtained \cite{AHM,Whit97} should
in fact be considered as
artifacts  coming from the finite-size effects
and/or inaccuracy in numerics.
Large oscillator strengths of the
{\em unconfined} $X^-_t$ ground state with $M_z < 0$ \cite{Chap}
is not compatible with the exact selection rule and is an error.
On the other hand, finite and large oscillator strengths of the
spatially confined $X^-_t$ ground state \cite{Hawr} (see also \cite{Chap})
results from the breaking of translational invariance.
The question now remains why in fact the $X^-_t$ ground state
is visible in experiment. One of the possibilities is localization
of charged excitons in real samples. This would
imply that the oscillator strength of the triplet ground state
should vary from sample to sample.
Another possibility comes from intrinsic effects and is associated with
the complex character of the valence band.
Finally, the rearrangement of the ground $X^-_t$
state with decreasing $B$ can in principle occur: with decreasing
separation between LL's the ground $X^-_t$ state could
jump from $M_z=-1$ to $M_z=0$. This would imply that
(1) there exist some critical field $B_c$ above which the $X^-_t$ state
abruptly becomes dark, and (2) the experimental situation
corresponds to $B<B_c$. Numerical calculations suggest that
the latter scenario appears not to be very plausible.
More theoretical work is needed to clarify the situation here.

We turn now to intraband transitions of charged magneto-excitons.
Such transitions in translationally invariant systems are
discussed theoretically in \cite{EP2DS},
where also experimental results for bound-to-continuum transitions are
reported and comparison between theory and experiment is made.
It was shown \cite{EP2DS,Dz&S_pr} that
the bound-to-bound transition from the triplet ground state
to a more stronger bound state associated with the next
\parbox{\columnwidth}{
\psfig{file=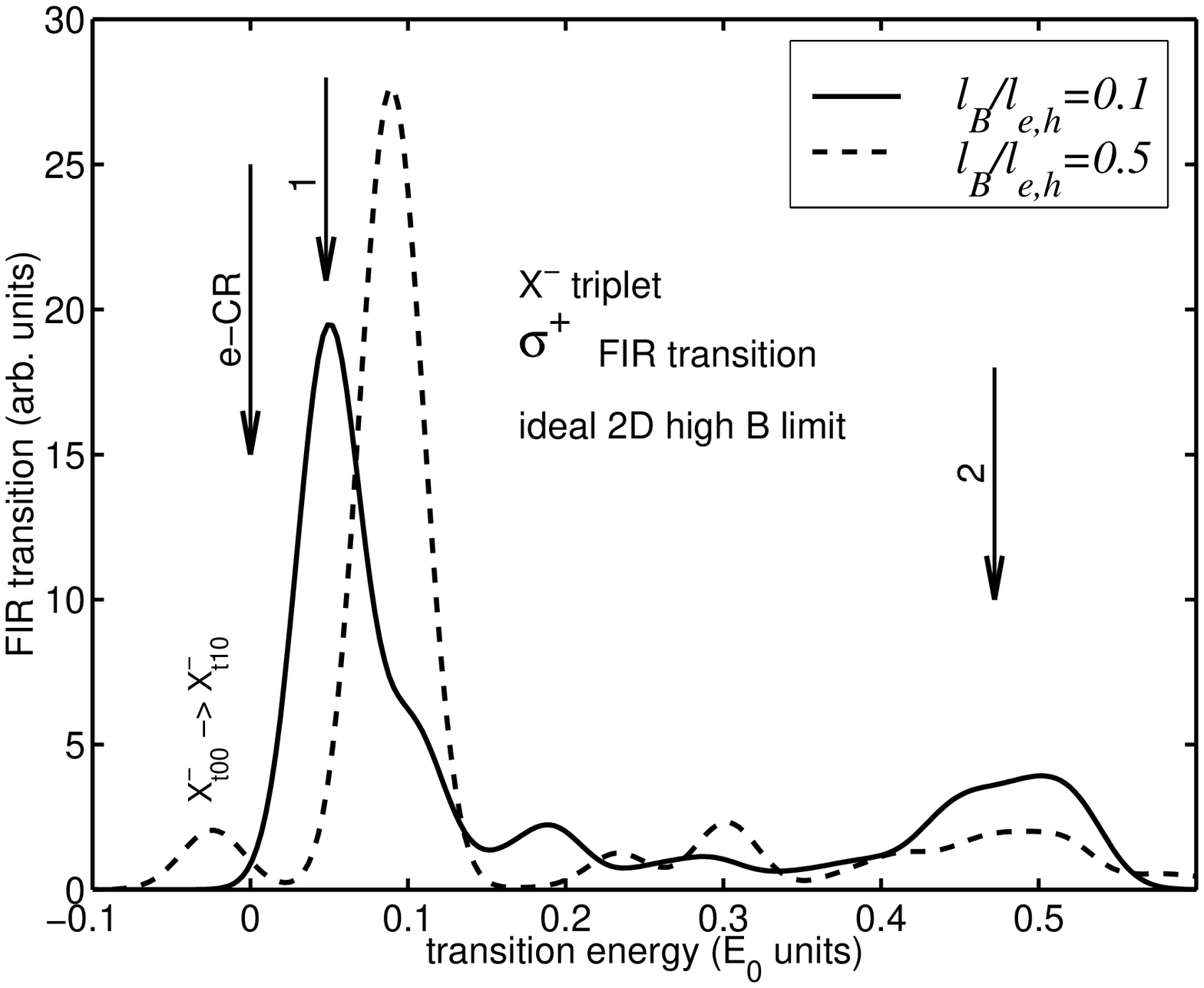,width=\columnwidth,height=7.5cm}
\small {\bf Fig. 1.}
Internal intraband transitions from the triplet $X^-_{t00}$ ground
state in the presence of parabolic lateral confinement
parametrized by lengths $l_e=l_h=l_{\rm e,h}$.
Strictly 2D system in the high-$B$ limit is considered.
Two cases of lateral confinement $l_B/l_{\rm e,h}=0.1$
and $l_B/l_{\rm e,h}=0.5$ are shown; $l_B=(\hbar c/ eB)^{1/2}$.
Energies are given in units of
$E_0=\protect\sqrt{\pi/2} \, e^2/\epsilon l_B$
relative to the confinement-renormalized $e$-CR energy shown by vertical arrow.
Vertical arrows 1 and 2 correspond, for the weak-confinement regime
$l_B/l_{\rm e,h}=0.1$,
to the bound-to-continuum transitions discussed in unconfined
systems in \protect\cite{EP2DS,Dz&S_pr}.

\vskip5mm
}
electron LL, $X_{t00} \rightarrow X_{t10}$, is prohibited.
Indeed the $X_{t00}$ PS has $M_z=-1$, while
the $X_{t10}$  PS has $M_z=1$, so that the usual selection rule
$\Delta M_z=\pm 1$ cannot be satisfied.
It is interesting to study how localization of charged excitons
breaks translational invariance and relaxes the $k$-conservation rule.
We model localization by the in-plane lateral spatial confinement,
which is assumed parabolic for electrons and holes
$V_e= \frac{1}{2} m_e\omega^2_e (x_e^2+y_e^2)$,
$V_h= \frac{1}{2} m_h\omega^2_h (x_h^2+y_h^2)$
with characteristic lengths
$l_e=(\hbar/2m_e\omega_e)^{1/2}$ and
$l_h=(\hbar/2m_h\omega_h)^{1/2}$, chosen such that
$l_e=l_h\equiv l_{\rm e,h}$
This model of confinement has been previously considered
for neutral \cite{Xdot} and charged \cite{Hawr,Chap} magnetoexcitons.
Here we present results for a strictly-2D situation in the
limit of high magnetic fields (no Landau level mixing).
The spectra of internal transitions from the triplet
ground state $X_{t00}$ to states in the next
electron LL for two regimes of lateral confinement are
shown in Fig.~1. Energies in Fig.~1 are given
relative to the confinement-renormalized electron
cyclotron-resonance ($e$-CR) $\hbar\tilde{\omega}_{\rm ce}  =
\hbar\omega_{\rm ce} \left( \sqrt{1+(l_B/l_{\rm e,h})^4} + 1 \right)/2$,
$l_B=(\hbar c/ eB)^{1/2}$.
In the regime of weak confinement $l_B/l_{\rm e,h}=0.1$ the
spectra practically reproduce the unconfined 2D case \cite{Dz&S_pr}.
For the intermediate confinement $l_B/l_{\rm e,h}=0.5$
the bound-to-bound $X^-_{t00} \protect\rightarrow X^-_{t10}$ transition,
which is prohibited in translationally-invariant systems,
develops {\em below} the $e$-CR. Development of such peak
is a tell-tale mark of localization of charged triplet
excitons.

In conclusion, we have studied the exact symmetry  ---
magnetic translations ---  for charged excitons in $B$ and
established its consequences for interband and intraband
optical transitions. In particular, we have shown
that in translationally invariant quasi-2D system with a simple valence
band the triplet ground state $X_t^-$ is dark in intraband transitions
at finite fields $B$. We have shown that in the presence of
symmetry-breaking effects the intraband bound-to-bound triplet
transition develops below the electron cyclotron resonance.
This suggests a method of studying localization of charged
excitons.

This work was financially supported in part by the DFG grant Os~98/5.
ABD is grateful to the Alexander von Humboldt Stiftung
for research support.


\end{multicols}

\end{document}